\documentclass[manuscript]{aastex}

\shorttitle{Recollimation shocks in 1803+784}
\shortauthors{Cawthorne et al.}

\begin{document}

\title{Polarization structure in the core of 1803+784: a signature of
recollimation shocks?}

\author{T. V. Cawthorne}
\affil{Jeremiah Horrocks Institute, University of Central Lancashire,
    Preston, Lancashire PR1 2HE, U.K.}

\and

\author{S. G. Jorstad\altaffilmark{1} and A. P. Marscher}
\affil{Institute for Astrophysical Research, Boston University, 725 Commonwealth Avenue, Boston, 
       MA 02215, U.S.A. }
\altaffiltext{1}{Astronomical Institute, St. Petersburg State University, Universitetsky
Pr. 28, Petrodvorets, 198504, St. Petersburg, Russia.}

\begin{abstract}
The extragalactic radio source 1803+784 is a BL~Lac object
that shows rapid variability and strong linear polarization. Very long baseline interferometry
observations reveal that the core
possesses a distinctive structure in polarized intensity with two maxima
on axis and two minima symmetrically placed on either side.
The approximately radial pattern of electric field polarization rods is reminiscent of the
results obtained earlier by Cawthorne (2006) for conical shocks, but, individually, these do not
reproduce the main features of the polarized intensity images. In numerical simulations
and experiments, these shocks occur in pairs and help to stabilize jets as they adjust
to changes in environment.
Here, the polarization resulting from such structures is investigated using an approximate,
analytical approach,
by making some simple assumptions about the nature of the flow
between two such shock waves.
For fairly small viewing angles, it is found that a reasonable representation of the
core polarization of 1803+784 can be obtained. The similarity between the observed and model polarization
supports the view that the core structure
in 1803+784 represents a recollimation shock, and that such shock waves may be responsible for the
first disturbance and hence brightening of the quiescent flow in astrophysical jets.
\end{abstract}


\section{Introduction}

When there is a mismatch between the pressure of a jet and the ambient pressure at the jet boundary,
the result can be a series of pairs of shocks  
that take the form of cone-shaped surfaces. In each pair of shock surfaces, a collimating or forward-pointing cone is 
followed by a decollimating or backward-pointing cone. It is likely that these two
shock fronts meet at a Mach disk, a small region over which the shock is planar and perpendicular 
to the axis. 

Since astrophysical jets are emerging into regions of declining pressure and density, it seems
probable that conical shocks are present and able to influence the observational properties of 
jets as they emerge from the active galactic nucleus. The effect of conical shocks on astrophysical
jets has been studied by a number of authors.  Lind and Blandford (1985)
considered their influence on the anisotropy of radiation from jets,
demonstrating that the result of relativistic beaming can 
differ quite significantly from that predicted by the usual relativistic beaming factor.  
Falle and Wilson (1985) performed 
numerical simulations of the jet flow in M87, identifying the conical shocks in their
simulations with the weak knots of radio emission between the nucleus and knot A. They found that,
using the rate of decline of gas pressure predicted by X-ray observations, their simulation
predicted the correct spacing of the knots.

G\'{o}mez et al. (1995, 1997) generated simulations of synchrotron emission from the output of hydrodynamic codes
in which recollimation shocks were produced by initially over-pressured jets in external media with both 
constant and declining pressure. G\'{o}mez et al. (1997) and Agudo et al. (2001) used this approach
to consider the interaction between moving shocks and
standing conical shocks, concluding that such interaction displaces the stationary component downstream and
gives rise to trailing components that follow the main outward-moving disturbance. 

Cawthorne and Cobb (1990) and Cawthorne (2006) used a simple semi-dynamical model for a conical shock front to
consider what the polarization signatures of such shocks might be. These models assumed that the
upstream magnetic field is either disordered on scales small compared to the resolution of any 
relevant observations or a combination of a disordered field component and a component of field parallel
to the axis, as is found in some astrophysical jets. The results were compared, for example, to
polarization observations of the $0.7$ arcsecond knot in 3C\,380 (Papageorgiou et al. 2006), which shows a 
pattern reminiscent of some of the simulations from Cawthorne (2006). Observations that revealed
a similar component in 3C\,120 have been reported by Roca-Sogorb et al. (2010) and Agudo et al. (2012). 
In the latter paper it was argued that this too represents a conical shock. {In 3C\,120, the case for 
the existence of a shock is strengthened by the large increase in surface brightness
over approximately $7$ years, which is interpreted as the the result of the formation
of the shock.

It is sometimes proposed (e.g., Daly \& Marscher, 1988, Marscher et al. 2008) that the high-frequency core components of astrophysical 
jets may be regarded as the first in the series of collimating and decollimating shock pairs that result 
when the
jet first becomes under-pressured. In this picture the jet is assumed to be essentially
invisible during its initial stages because it is a steady flow and therefore `cool'.
Such shocks seem to be an essential feature of most AGN jets because, unless a jet evolves smoothly
(which is not, apparently, the case for most jets, which contain bright and variable features) such shocks represent the only
mechanism through which it can reach a form of equilibrium following changes in the pressure of the jet or of its environment. 
Indirect evidence for identification of the high-frequency radio cores with recollimation shocks comes from the work of 
Marscher et al. (2008) 
who attribute flares in a some AGN to the interaction of a travelling shock with a stationary recollimation shock at the
position of the core.   
However, in these circumstances, where the features concerned are at best marginally resolved by VLBI arrays, 
key signatures of the shocks, such as the ratio of the upstream to downstream brightness, cannot be measured directly or
even inferred (as in the case of the more extended structure in 3C\,120). For this reason, the best way to confirm
the existence of recollimation shocks is to 
identify their characteristic features, among which the polarization signature is one of the most promising candidates.

In the past, comparison between model and observed polarization
has been difficult since, at the resolution of available data, the core
represents a blend of many different components and is often dominated by newly emerging knots rather
than the quiescent features that might be associated with the conical shocks. Recently, however, 
the availability of high quality images from the Very Long Baseline Array (VLBA) at frequency $43$\,GHz has provided a 
new source of data in which to search for radio cores that might consist of simple, quiescent
structures. Despite the availability of such images for a number of years,
the polarization is often not displayed optimally for this purpose. Visualizations in which the polarized intensity
is represented by the length of polarization rods often give a poor impression of its distribution in marginally resolved
structures
- and are not ideal for comparison with models. Contour maps 
of polarized flux density with polarization rods superimposed
are far better in this respect, but rarely seem to appear in published work. 

This paper presents images of the quasar 1803+784 drawn from the $7$\,mm VLBA polarization 
survey by Jorstad et al. (2005, 2007), the polarization contour maps being shown here for the first time. 
They reveal a distinctive polarization structure that is discussed in terms of the conical shock models. 
 
\section {Observations}
The compact, flat-spectrum radio source $1803+784$ is associated with a BL~Lac object at redshift 
$z=0.6797$ (Lawrence et al., 1996). The structure of the radio source takes the form of a 
one-sided and apparently curved jet. A series of bimonthly VLBI observations at $43$\,GHz revealed three 
superluminal components with apparent speeds in the range $\simeq 10-16\,c$ (Jorstad et al., 2005).
Observations at lower frequencies with less frequent sampling suggested that most prominent 
features in the jet were
stationary, though one superluminal component with apparent speed $\simeq 19c$ was detected
(Britzen et al. 2010). 

The observations used in this paper were obtained as part of a larger program of
investigation in which a sample of AGN was observed at roughly 2 month intervals
over the course of approximately 3 years, from 1998 to 2001. The sources were observed by the VLBA 
at frequency
$43$\,GHz  and by a number of other telescopes operating in the mm, sub-mm and optical
bands. The full results were presented by Jorstad et al. (2005) and Jorstad et al. (2007). 

\begin{figure*}
\begin{center}
\includegraphics[angle=0,width=13.75cm]
{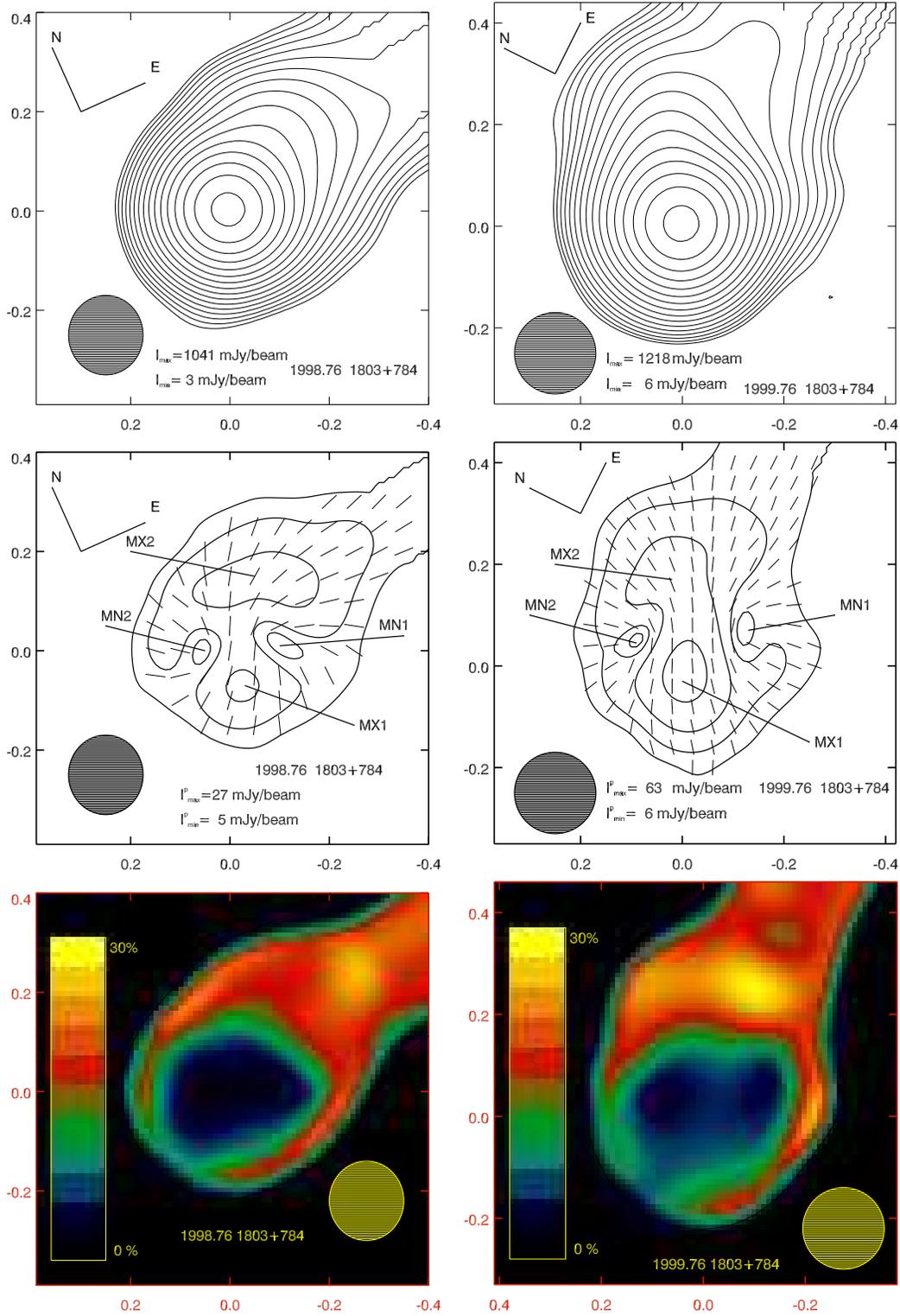}
\caption {Images of 1803+784 at $43$\,GHz in total intensity, polarized intensity and 
polarization $E$-field rods of constant length, and fractional polarization. The total intensity contours
increment by a factor of $\sqrt{2}$ and the polarized intensity contours increment by a factor of $2$. The 
RA and Dec scale is in mas. The polarization minima and maxima are labelled MN1 and MN2, and MX1 and MX2, respectively.}
\end{center}
\end{figure*}

Inspection of images of 1803+784 presented in Jorstad et al. (2005) reveals radial patterns of 
polarization rods, reminiscent of results from the conical shock modelling of Cawthorne (2006). To facilitate 
a closer comparison, the polarized intensity contour images, in which the polarization angles are 
represented by rods of constant length, were generated, and two of these are shown in Fig.\,1, where 
total intensity contour images and fractional polarization gray-scale images are also shown. 
The images have been rotated so that the apparent jet axes are vertical. The direction of North is
shown in the total and polarized intensity plots.

In this paper, Faraday rotation is assumed to be unimportant at $43$\,GHz in 1803+784. This is implied by the 
approximate
reflection symmetry of the polarization rods about the jet axis, as seen in Fig.\,1; it is unlikely that 
such symmetry would survive 
a significant amount of Faraday rotation. Furthermore, Mahmud et al. (2009) measured Faraday rotation 
associated with
core region, using images at $43$\,GHz and lower frequencies. The largest RM detected was $1200$\,rad\,m$^{-2}$, 
corresponding to a rotation of only $3.5^{\circ}$ at $43$\,GHz.

The most striking aspect of these images is the unusual and complex 
polarized intensity structure.
As expected, many of the images do show a radial pattern of polarization rods similar 
to that found in some of the conical shock-simulations shown by Cawthorne (2006). However there are also
features that the conical shock models do not reproduce. The polarized intensity is often 
double-peaked. These two peaks, which define an approximate symmetry axis, are of comparable 
brightness, the North-Western of the two (i.e. the furthermost downstream) being weaker at some
epochs and generally less compact. 
On a line orthogonal to the symmetry axis between the two peaks, there are two minima of polarization,
one either side of the axis. This type of structure is particularly evident at the epochs October 1998 and October 1999 
shown in Fig.\,1 but is present to some extent at most of the seventeen observations that were made. 
For clarity, the maxima and minima of polarized intensity are marked on the corresponding plots in Fig.\,1.

This paper addresses the question of whether such structures could arise from the 
conical shock model or one of its variants. 

\section {Models}

Any model for the structures shown in Fig.\,1 must account for the two minima in polarized intensity, and
must therefore entail cancellation between orthogonal components of polarization.
In addition, the model must explain the two on-axis components of polarized intensity and 
pattern of polarization rods that radiate from a position between them. 
The required cancellation of polarization can result in a number of ways. For example, if the jet
contained a partially ordered magnetic field, then such an effect might result from
superposition of emission from toroidal and poloidal field components;
such models are discussed, e.g., in Canvin and Laing (2004). 
However, the radial pattern of polarization rods is reminiscent of the conical shock models of Cawthorne \& Cobb (1990)
and Cawthorne (2006). These form an attractive basis for this investigation, since, as discussed in Section 1,
they are a component of recollimation
shocks, which are often identified with the high frequency cores in AGN (e.g., Marscher et al., 2008). 
In this model the polarization minima result from superposition of polarized emission from the collimating and decollimating
shocks: in the region between the polarized maxima,
their polarization angles rotate in opposite directions with increasing distance from the axis,
making this type of cancellation a very probable occurrence.

\subsection{Conical shock models.}

The models used in this paper are essentially those of Cawthorne (2006), based on a framework presented in
Lind \& Blandford (1985).
The shocks are modelled as conical 
surfaces, having semi-opening
angle $\eta$, that are coaxial with the jet. 
The axis of the system is inclined at an angle $\theta$ to the line of sight, where $\theta$ 
and $\eta$ are measured in the reference frame of
the shock (which is assumed to be at rest with respect to the AGN). In the region upstream of 
shock the flow is assumed to
be parallel to the jet axis with speed $c\beta_u$. As a result of the first shock, the flow is deflected 
through an angle $\xi$ away from the
axis and decelerated to speed $c\beta_d$. The system is shown in Fig.\,2.

\begin{figure}
\begin{center}
\includegraphics[angle=0,width=8cm]
{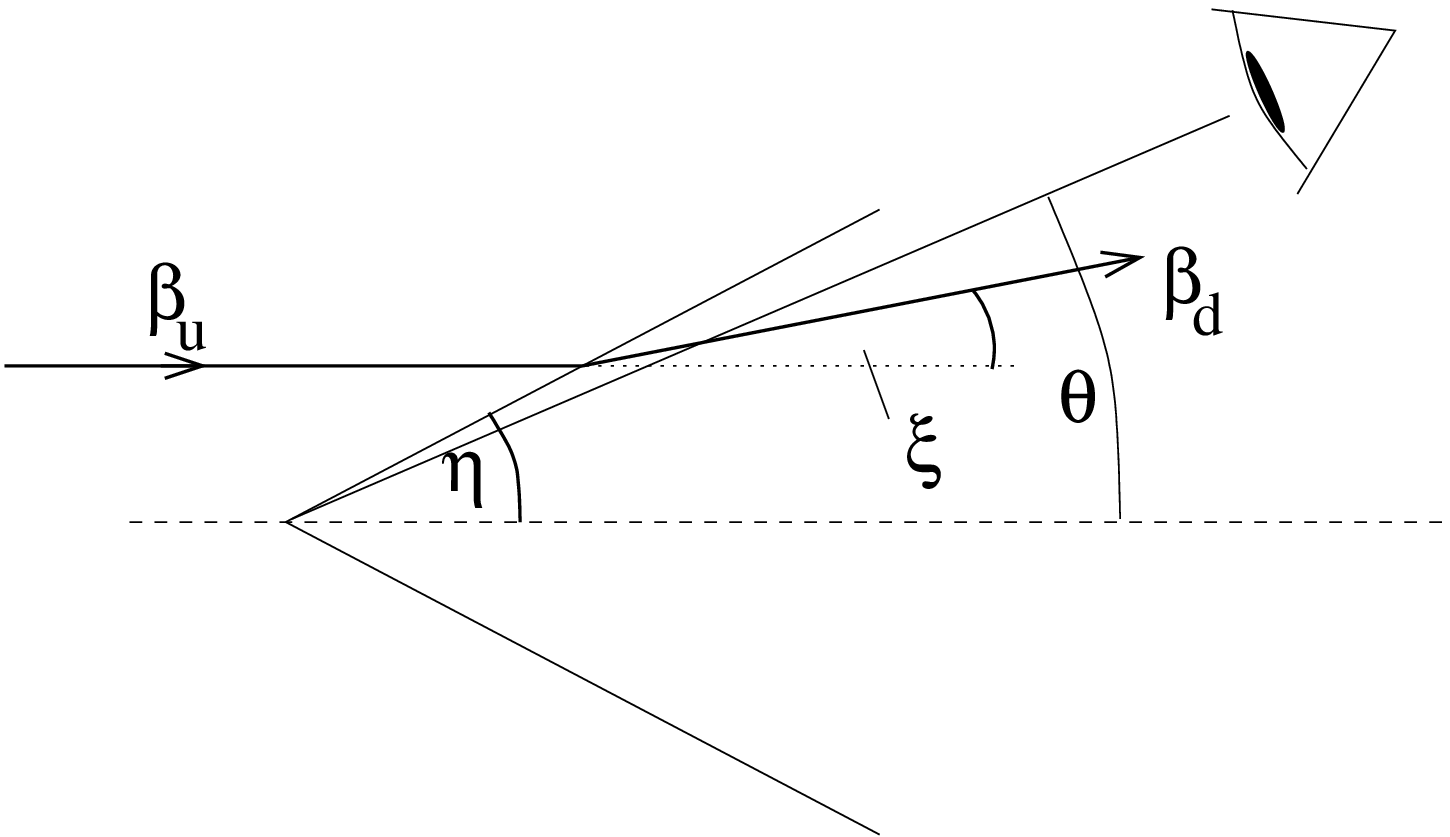}
\caption {Illustration of conical model. The flow is initially parallel
to the jet axis, but is deflected by the shock through an angle $\xi$.}
\end{center}
\end{figure}

\begin{figure*}
\begin{center}
\includegraphics[angle=-90,width=15cm]
{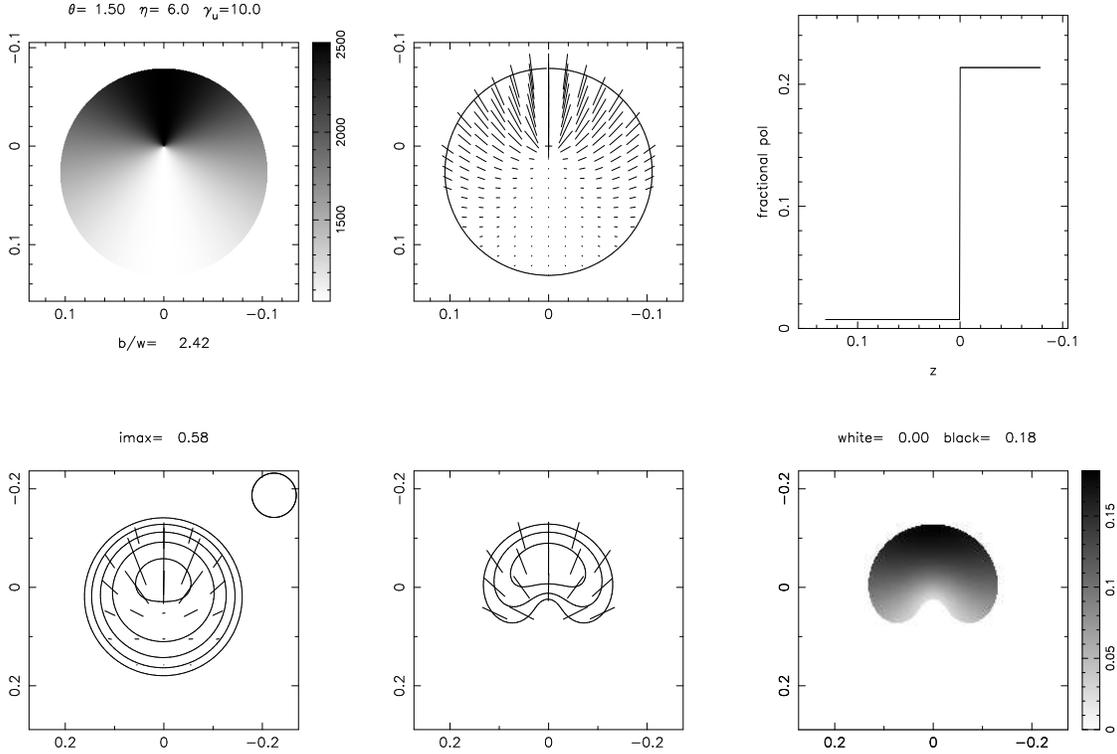}
\caption {Illustration of emission from the single conical shock model with upstream Lorentz factor 
$\gamma_u=10$, cone semi-angle
$\eta=6^{\circ}$, and an angle $\theta=1.5^{\circ}$ between the axis and line of sight. 
The first two panels show distributions of
total intensity, $I$, and polarized intensity, $P$. The third panel shows the run of fractional polarization 
along the axis.
The fourth and fifth panels illustrate the effects of convolution with the circular Gaussian beam shown in the 
top right of the 
fourth panel. They show, respectively, contours of $I$ and $P$, 
with polarization $E$ field rods proportional in length to $P$ (fourth panel) 
and constant in length (fifth panel). The value of $I_{max}$ shown above the fourth panel represents the length
of a polarization rod that corresponds to an intensity equal to the peak value of $I$. The lowest contours are 
$5\%$ in $I$ and $15\%$ in $P$, with contours
stepping upward in factors of two. The sixth panel shows the distribution of fractional polarization post
convolution, cut at the lowest $P$ contour.
}
\end{center}
\end{figure*}

In previous work (Cawthorne \& Cobb 1990, Cawthorne 2006, Agudo et al. 2012) simulated polarization images
were formed by making  
the following assumptions: the emission is 
dominated by one of the two conical shock waves
(or, in the case of Agudo et al. 2012, only one such shock occurs);
the shock structures are assumed to be steady over timescales less than or comparable to
a light crossing time; the source is assumed to be optically thin for frequencies at which the emission
is to be simulated;
the radiating material is visible only near the shock
surface, i.e., it cools rapidly; and the upstream densities of radiating particles 
and magnetic energy are uniform across the jet's cross section. Both up and downstream
plasmas are assumed to have a relativistic equation of state and sound speed $c/\sqrt{3}$.

Among models of this type, those of greatest interest here have jet axes making small angles to 
the line of sight and largely disordered upstream magnetic fields (corresponding to $f=0$ in Cawthorne, 2006). 
It is these models that
reproduce the roughly radial pattern of $E$-field polarization rods found in the core of 1803+784. 

An example is shown in Fig. 3. Here, the emission is computed for a single conical shock front where 
the jet axis is inclined at angle $\theta=1.5^{\circ}$ to 
the line of sight, the upstream Lorentz factor is $\gamma_1=10$ and the cone angle is $\eta=6^{\circ}$.
With these values, the downstream Lorentz factor is $\gamma_2=7.8$, and the 
flow is deflected toward the shock surface by $2.1^{\circ}$. 

The first and second panels show model total and polarized intensities, 
respectively. The third panel shows variation of fractional polarization along the axis. The fourth,
fifth and sixth panels show the effects of convolution with a beam: the fourth panel shows total intensity
contours with polarization rods of length proportional to polarized intensity and the restoring beam (FWHM) in the top right;
the fifth panel shows polarized intensity contours 
with polarization rods of constant length; the sixth panel shows a gray-scale plot of fractional polarization.

Even when the jet axis makes such a small angle to the  line of sight, the 
intensity patterns are still quite strongly asymmetric, in the sense that the total intensity and
polarization are higher on the 
downstream (upper) than on the upstream (lower) side of the shock.
In total intensity, this
arises because the angle between the downstream velocity and the line
of sight is much smaller, and therefore the Doppler factor is much larger,
on the upper than on the lower part of the cone surface. 
The same
effect also boosts the polarized intensity, but the main reason for the 
polarization asymmetry is that,
in the rest frame of the downstream flow, the angle between the line of sight and the normal to shock surface
is much closer to $90^{\circ}$ on the upper than on the lower part of the cone surface. Hence, the
magnetic field appears much more highly ordered on the upper shock surface, and the fractional polarization
is therefore much higher there than on the lower side. This is clearly seen in panel 3 of Fig.\,3.

Note that in these simulations, the shock can be regarded either as decollimating (in which case the flow is  
downward in the figure and the conical surface is opening toward the observer) or collimating (when the flow is
upward and the conical surface opens away from the observer) (see also Fig.\,2 in Cawthorne, 2006).

While the polarization distribution does reproduce approximately a radial pattern of polarization,
it reproduces neither the double peak in polarized intensity seen on the axis of the core in 1803+784, nor the 
polarization minima either side of it.   
If the viewing angle is increased, the polarization becomes rapidly more strongly asymmetric, leaving just a
simple fan of polarization rods associated with the upper part of the shock.
As the viewing angle is reduced further, so the entire polarization structure rapidly 
becomes circularly symmetric. It seems that, with a purely disordered upstream magnetic field, there is
no combination of parameters that will reproduce the polarization distributions shown in Fig.\,1. 
At this point, therefore, an extension of the conical shock model is considered.

\subsection{Recollimation shocks.}

In numerical simulations and laboratory experiments, conical shocks occur in pairs. 
The first, collimating shock 
over-compensates for the initial divergence of the flow, and this is corrected by the second, decollimating shock.
The two shocks are coaxial and meet at their apexes, or possibly at a small Mach disc.
In this paper, such structures are represented by the model shown in Fig.\,4.
The polarization distribution from
such a system would be the sum of two distributions, each similar to that of Fig.\,3, panel 5, 
one being reflected about 
the horizontal axis of the plot. This clearly offers the possibility of a structure with two 
peaks in polarized flux density on the jet axis. 
Slightly less obviously, it also raises the possibility of pairs of polarization minima along a line perpendicular 
to the axis.
These minima result from cancellation due to the almost
orthogonal polarizations contributed by the convolved images of the two shock structures. It therefore seems 
worth exploring whether the observed 
polarization could result from this type of structure. 

Using the analytical approach,
the polarization of synchrotron radiation from such structures can only be determined by introducing some additional
assumptions, above and beyond those used by Cawthorne (2006), most of which relate to the behavior of the plasma 
between the two shocks; they are:
\begin{enumerate}
\item The initial and final fluid velocities are essentially parallel to the jet axis. In fact, numerical
simulations show that in both regions, the flow is likely to be parallel to the axis near the axis, but diverging
away from the axis at larger radii (J.L. Gomez, private communication). This
would be difficult to combine with the approach used here, where the requirement that the shock front is a perfect
cone demands that the flow velocities (within any half-cross section) are parallel.
\item The flow is diverted toward the axis by the first, collimating shock, and then realigned with the axis by the 
second (decollimating) shock. The cone angle of the second shock is determined by the need to realign the flow.
(The effects of the converging upstream flow at the second shock can be taken into account using the approach of 
Nalewajko, 2009).
\item It is assumed that, downstream from the first shock, the plasma is made turbulent, such that following its
passage through the emitting layer, the magnetic field returns to a disordered state. 
The high radio frequency core polarization in AGN is generally low and also highly variable both in intensity and 
position angle, a fact 
that is best explained if the upstream magnetic field is largely turbulent (e.g. Hughes, 2005).
This is also true of the optical polarization, and the correlation between the polarization angles in the two frequency bands
suggests the emission arises in the same region (Gabuzda \& Sitko 1994, D'Arcangelo et al., 2007, D'Arcangelo, 2010). 
In the case of
0420--014, D'Arcangelo et al. (2007) estimated that the emitting region consisted of some $600$ cells, so the scale
of the cells is likely to be much smaller than the size of the emitting region. 
Shocks are likely to cause a turbulent downstream flow if the gas pressure exceeds the magnetic pressure (an implicit
assumption here, since the effect of the field on the shock dynamics has been ignored) (e.g., Hughes \& Eilek, 1990).
If the downstream flow is turbulent on a similar length scale the to the upstream flow, then the imprint of the
shock on the magnetic field structure could be lost quite quickly, even if the turbulent velocity component is significantly 
less than the flow velocity. Observationally,
there is relatively little evidence to show how far downstream the ordered field survives, due to the difficulty of imaging
interknot emission.
\item As in Cawthorne (2006), the emitting
plasma is assumed to cool rapidly. Here, it is further assumed that the rest-frame plasma emissivity returns 
rapidly to its upstream value, and that emitting layers at the two shock fronts are of equal width.
The most likely cooling process is synchrotron radiation, though inverse Compton scattering of synchrotron radiation and
of high frequency (optical or infra-red) radiation from an accretion torus may also contribute. The validity of this assumption
is discussed at the end of this section.
Note, however, that if the electrons do age rapidly through radiative cooling,
and if the core component could be isolated from its neighbouring features, 
spectral ageing should be manifest in a steepening of the spectrum at frequencies above $40$\,GHz. This could be an
important test of the model when suitable spectra are available.

\item Between the two shocks, the flow is converging and the plasma is subject to a two--dimensional, radial
compression. During this process, it is assumed that the disordered state of the magnetic field is maintained by
turbulence. 
\item This compression causes an enhancement of the magnetic field and the particle density, potentially by a large factor.
If the re-tangling of the field does not increase the magnetic flux density (as it may well, e.g. Scheuer, 1987), 
then compression of the jet radius by a factor $(r/r_0)$ may result in an enhancement in synchrotron emissivity by a 
factor as 
great as $(r/r_0)^{-q}$ where $q=4\frac{2}{3}$ or $q=6$ for spectral index $\alpha=0.5$ or $1$, respectively (Leahy, 1991). 
Such a strong dependence presents a problem for the model proposed here, as it results in 
images that are completely dominated by the second shock, disallowing the formation of polarized features such 
as double peaks on axis and polarization minima.
The effect of finite synchrotron lifetime may mitigate the effect of compression. However,
the root of the problem is that the semi-analytical model
of Fig.\,4 almost certainly produces too great a convergence of the flow between the two shocks, as it neglects any
response of the flow to increasing internal pressure. 
This view is supported by
the MHD simulations produced by Roca-Sogorb et al. (2008) and Roca-Sogorb (2011) which clearly indicate recollimation shocks 
in which the largest width of the decollimating shock is not much smaller than that of the collimating shock. 
This is clearly a failing of the semi-analytic model. 
As a result of this, and in the absence of any obvious analytical way to allow for effects that might moderate
the sensitivity of the plasma emissivity, slightly weaker (and ad hoc) forms of dependence have been investigated. 
It was found that an emissivity proportional to $(r/r_0)^{-n}$ where $n \simeq 3$
allowed some interaction between the polarizations of the two shocks, and 
this dependence has been adopted for the purpose of the present work. 
This ad hoc solution is, of course, not entirely satisfactory, but it provides a convenient way to proceed, pending
a fully--numerical investigation.
\end{enumerate}

\begin{figure}
\begin{center}
\includegraphics[angle=0,width=10.5cm]
{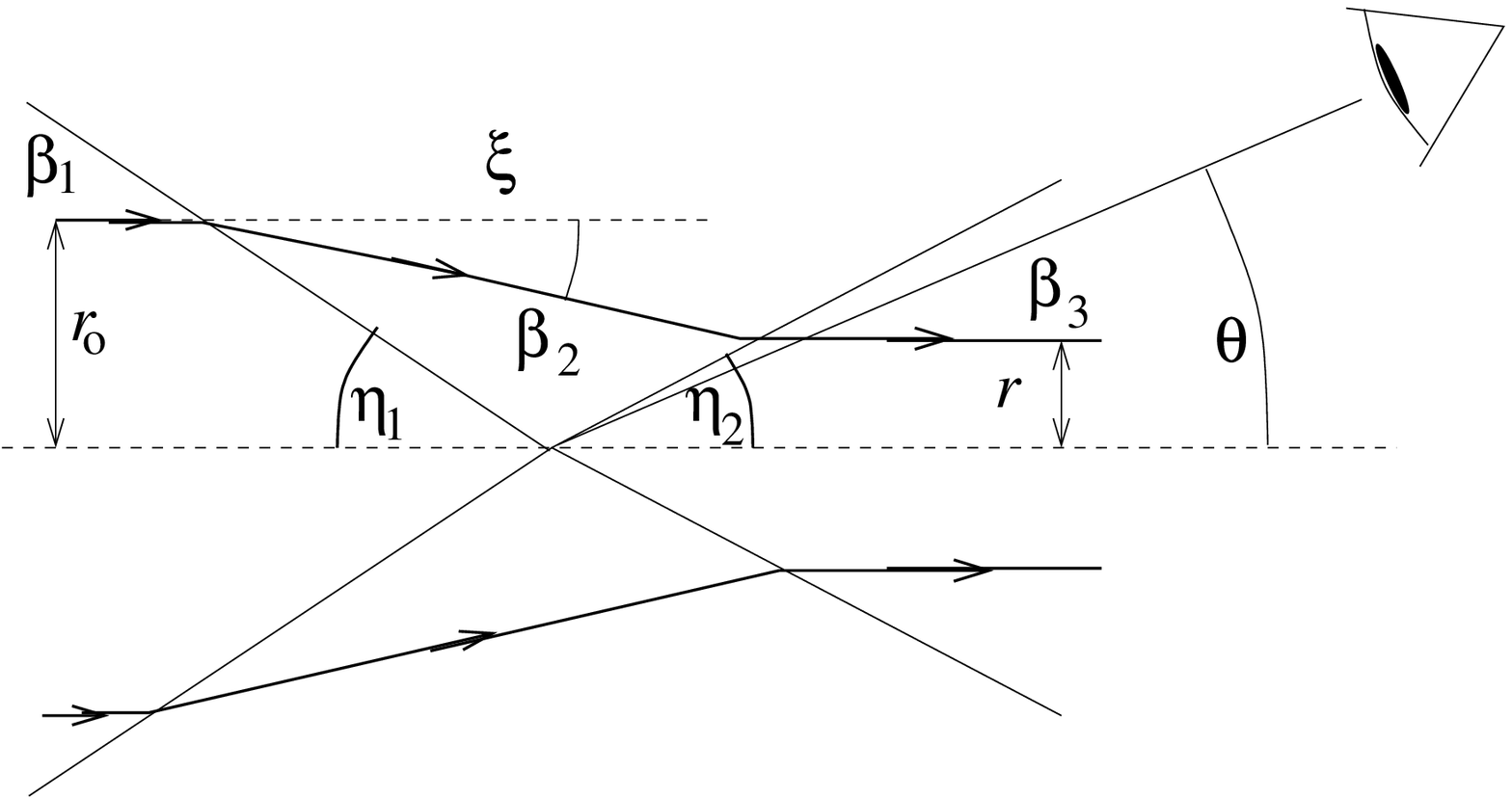}
\caption {Illustration of recollimation shock model. The flow is initially parallel
to the jet axis, but is deflected by the first shock through an angle $\xi$ then realigned with the axis by
the second shock.}
\end{center}
\end{figure}

\begin{figure*}
\begin{center}
\includegraphics[angle=-90,width=16.5cm]{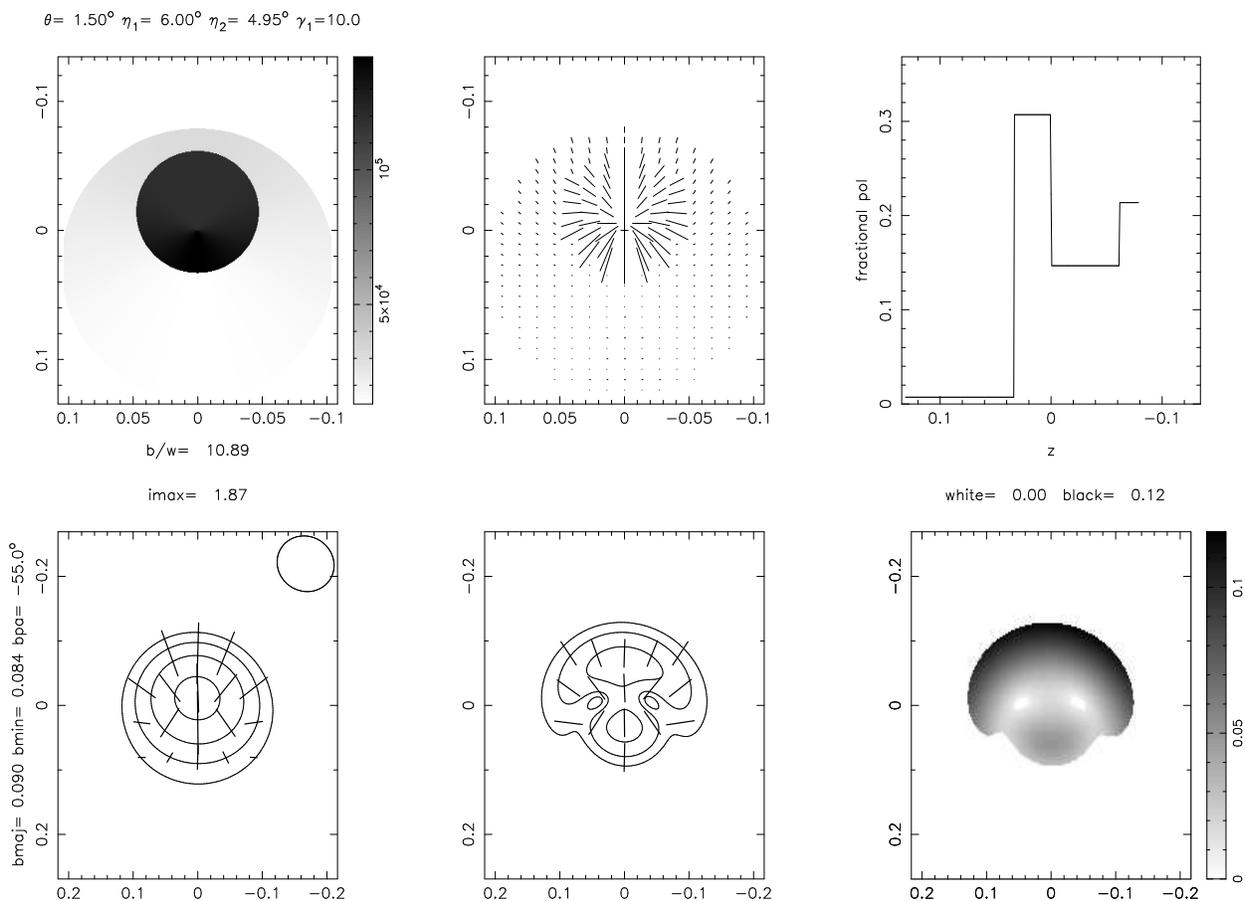}
\caption{Images for the double shock model corresponding to parameter values shown above the first panel. 
The panels are as described in Fig.\,3. The lowest contours in the convolved total and polarized intensity
contour maps are $10$ and $18$ percent of their respective peak values.
}

\end{center}
\end{figure*}

\begin{table*}
\begin{center}
\caption{Comparison between data and model parameters for October 1998 observations.}
\begin{tabular}{llllll}
\tableline\tableline
     & $\Delta r_{max}/b_{maj}$ & $\Delta r_{min}/b_{maj}$ & $p_{max}/p_{min}$ & $m_{core}$ & $m_{jet}$   \\
\tableline
data & $1.38 \pm 0.13$          & $1.00 \pm 0.088$ & $1.06 \pm 0.18$     & $0.032 \pm 0.004$ & $0.084 \pm 0.008$  \\
model& $1.05$                   & $0.97$           & $1.09$              & $0.038$           &   $0.049$       \\
$|{\rm data}-{\rm model}|/\sigma$
     & $2.6$                    & $0.4$            & $0.2$               & $1.6$             &  $4.4$   \\
\tableline
\end{tabular}
\end{center}
\end{table*}

\begin{table*}
\begin{center}
\caption{Comparison between data and model parameters for October 1999 observations.}
\begin{tabular}{llllll}
\tableline\tableline
     & $\Delta r_{max}/b_{maj}$ & $\Delta r_{min}/b_{maj}$ & $p_{max}/p_{min}$ & $m_{core}$      & $m_{jet}$   \\
\tableline
data & $1.25 \pm 0.25$          & $1.30 \pm 0.13$         & $0.48 \pm 0.09$   & $0.05 \pm 0.005$ & $0.15 \pm 0.03$  \\
model& $1.06$                   & $1.11$                  & $0.72$            & $0.06$           & $0.04$       \\
$|{\rm data}-{\rm model}|/\sigma$
     & $0.75$                    & $1.5$                  & $2.6$             & $2.0$            &  $3.6$ \\  
\tableline
\end{tabular}
\end{center}
\end{table*}

These assumptions are required to connect the states of
the fluid at the two shock fronts. 
They allow this investigation to determine whether
the idea of using twin recollimation shocks to explain the core polarization of 1803+784 is a plausible one. 
If the results are promising, then the behavior of the plasma between the shocks needs to be investigated using 
a numerical approach. 

The geometry of the shock structure is summarized in Fig.\,4. 
This model is characterized by three input parameters, namely, the upstream Lorentz factor $\gamma_1$, the cone angle of
the first (decollimating) shock, $\eta_1$, and the viewing angle $\theta$. That said, there is an approximate equivalence 
of simulated images for which $1/\gamma_1$, $\eta_1$, and $\theta$ are scaled in proportion; it is thus the two parameters
$\theta/\gamma_1^{-1}$ and $\eta/\gamma_1^{-1}$ that determine the results.
In addition, for any value of
$\gamma_1$, the range of values of $\eta$ is constrained fairly strongly: $\eta$ must exceed the minimum value for
the formation of a shock front, given by $\sin(\eta_{min})=1/(\sqrt{2}\gamma\beta)$; however, values much larger than
this result in large deflections, strong compression of the plasma before it reaches the second shock, and significant 
reductions in the Lorentz factor of the flow, none of which is helpful in explaining the structure of 1803+784. 
Furthermore, the value of 
the viewing angle, $\theta$, 
is constrained to be less than the cone angle (to obtain the required radial pattern of polarization $E$ field). 
Hence the range of model parameters to be investigated is quite tightly
restricted. As a result of the first shock the flow is deflected through an angle $\xi$ toward the axis and decelerated to
speed $c\beta_2$. The second shock realigns the flow with the jet axis, further reducing the flow speed to $c\beta_3$. 

The choice of Lorentz factor for 1803+784 is not straightforward, as there are several superluminal components with 
a range of speeds (which themselves, vary with time)
(Jorstad et al. 2005 and Britzen et al. 2010). Fortunately, as discussed above, this choice is not critical and for 
the present work
a nominal upstream Lorentz factor $\gamma_1 = 10$ has been adopted. Doubling the Lorentz factor and halving the angles
$\theta$ and $\eta$ (and the beam size) produces relatively minor changes in the simulated images.

The smallest possible choice of angle $\eta_1$,  for the
upstream shock, is $\eta_{min} \simeq 4.1^{\circ}$, while for $\eta_1=7^{\circ}$, the jet Lorentz factor is halved 
on passage through
the two shocks, and the deflection angle of $\simeq 3^{\circ}$ results in halving of the flow radius by the time the second
shock is encountered. The results of greatest interest are therefore to be found between these two values of $\eta_1$. 

In an attempt to characterize the degree of success of the simulations, several characteristic properties of the 
polarized intensity images were
measured. 
These are: $\Delta r_{max}/b_{maj}$, the ratio of the separation of the polarization
peaks on axis to the FWHM major axis of the beam, $b_{maj}$; $\Delta r_{min}/b_{maj}$,  
the ratio of the separation of the polarization
minima to $b_{maj}$; $p_{core}/p_{jet}$, the ratio of the flux densities of the two polarized peaks, and 
$m_{core}$ and $m_{jet}$,
the fractional polarizations of these two features. (In this context, the upstream and downstream peaks in polarized
intensity are labelled, respectively, as the `core' and `jet' components without implying any physical connotation.)
The values of these quantities were obtained by plotting profiles 
of total and
polarized intensity and fractional polarization along lines running along the axis, and perpendicular to the axis 
along a line
plotted through the two minima. The uncertainties were also estimated from 
these plots, those
in polarized flux density and fractional polarization being obtained as described by Hovatta et al. (2012).  A $\chi^2$ 
parameter
was defined for each of these quantities and the total $\chi^2$ was used to judge the level of agreement between model 
and data. A range of models was explored manually, by varying
$\eta$ and $\theta$ in $0.25^{\circ}$ intervals. 
The size of the beam is, of course, unknown in model coordinates, 
For each
pair of values $\eta$,\,$\theta$, the major axis was varied to find the value that minimized $\chi^2$. 
(The ratio $b_{maj}/b_{min}$
and the beam position angle relative to the jet axis were held at the known values of $1.07$ and $55^{\circ}$, 
respectively.) The comparison between the observed and model values of $\Delta r_{max}/b_{maj}$ and
$\Delta r_{min}/b_{maj}$ provides a check for the self--consistency of this procedure.

It was found that, for $\gamma_1=10$ and values of $\eta_1$ in the region of greatest interest, 
$4.1^{\circ}<\eta_1<7^{\circ}$,
the main features of the observations, i.e., the double peaks on axis and the double minima of the polarized intensity, 
were obtained for
values of $\theta$ in the approximate range $1^{\circ}<\theta <2^{\circ}$. 
The values of the characteristic parameters and the deviations between model and observed values
are shown in  Tables\,1 and 2 for the October 1998 and October 1999 images, respectively. 
For the October 1998 observations shown in Fig.\,1,
at $\eta_1=6^{\circ}$, the smallest $\chi^2$
occurs for $\theta=1.5^{\circ}$.
The value of $\eta_2$ for this model is $4.95^{\circ}$ and the deflection angle $\xi=2.14^{\circ}$.  
The simulated images for this model are shown in Fig.\,5, in which the six panels are as described for Fig.\,3.
To facilitate comparison of the observed and simulated polarized intensity images, these are shown side by side in 
Fig.\,6.

\begin{figure}
\begin{center}
\includegraphics[angle=0,width=12cm]{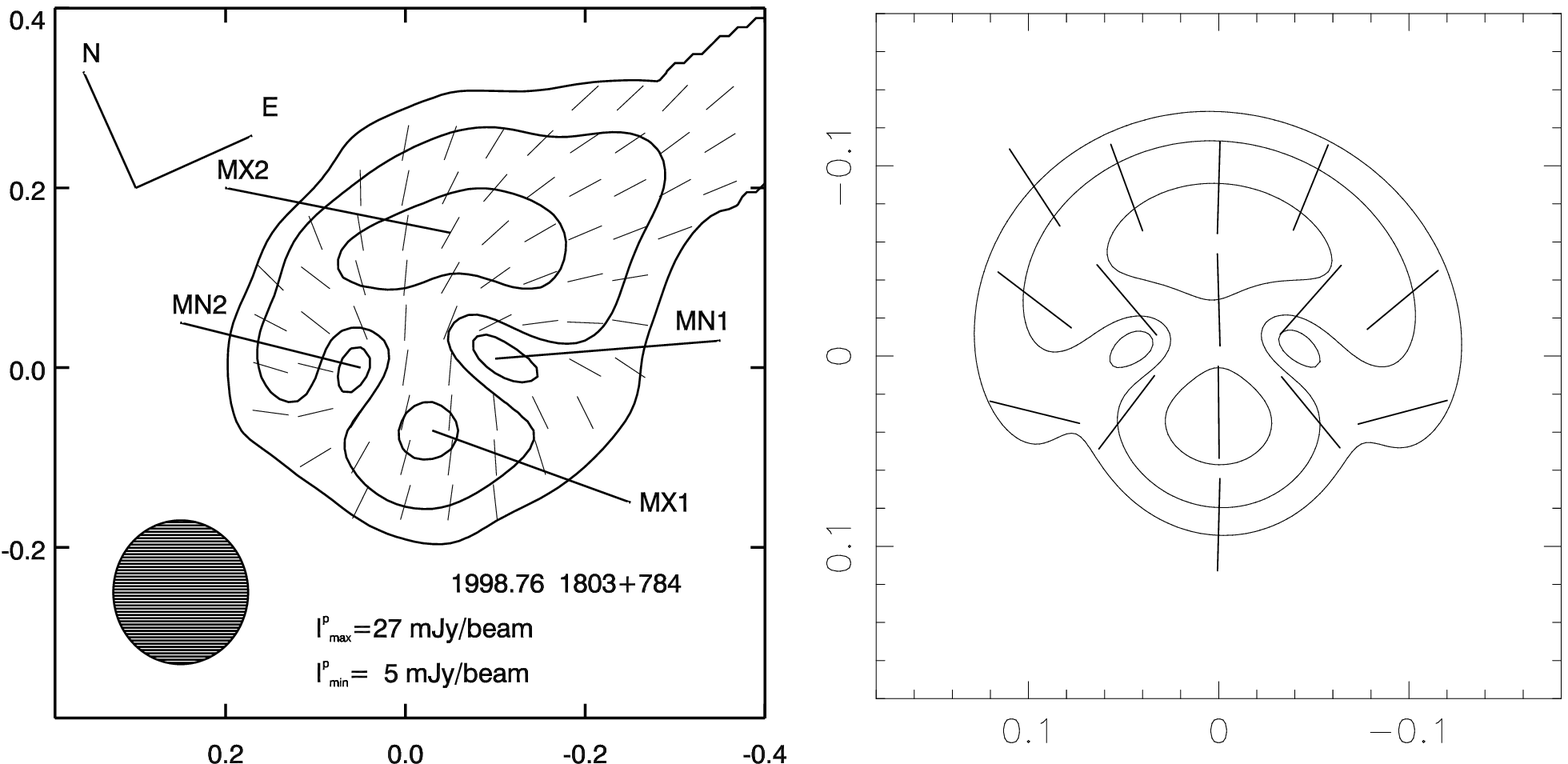}
\caption{The October 1998 polarized intensity image from  Fig.\,1 is shown alongside the simulated polarized intensity image 
from Fig.\,5, to facilitate comparison of the two. 
}
\end{center}
\end{figure}

For the October 1999 observations, also shown in Fig.\,1, a slightly smaller cone angle is favored. With
$\eta_1=5^{\circ}$, $\eta_2=4.45^{\circ}$ and $\xi=1.14^{\circ}$, the smallest $\chi^2$ occurs for $\theta=1.75^{\circ}$. 
The simulated image for this model is
similar to that shown in Fig.\,5 (except that the `jet' polarized intensity is less than that of the `core') 
and so is not presented separately. 

Formally, the fits are not good.
For both observations, the major source of discrepancy is the polarization of the outer `jet' component, which
has a much higher polarization than can be obtained in this model. 

At this point,  it is important to consider whether Assumption 4, that shocked gas cools 
quickly, is valid. The synchrotron lifetime of electrons in the shocked gas is
\begin{eqnarray}
\tau_{synch} &\simeq& 1.5 (B/5 \times 10^{-5}\rm{T})^{-3/2}(\nu_{obs}/43\,GHz)^{-1/2} ~\rm{yr} \label{tsyn}
\end{eqnarray}
where the emission and observing frequencies are related by $\nu_{em}=\nu_{obs}(1+z)/D$, and
a typical Doppler factor $D \simeq 13$ for the October 1998 model has been assumed.  
Given that, on submilliarcsecond scales, values several times $B\simeq 10^{-5}$\,T are quite probable (e.g., O'Sullivan \& Gabuzda, 2009),
synchrotron lifetimes may be as short as $1.5$ years, resulting in a downstream flow distance 
$\gamma_2 v_2 \tau \simeq 10.5$\,ly\,=\,3.2\,pc (for $\gamma_2=7$). This compares to a deprojected separation of the 
two polarized maxima of
$\simeq 60$\,pc (for an inclination angle of $1.5^{\circ}$). Therefore, so long as the magnetic field is greater than about 
$10^{-5}$\,T, the emitting gas will
cool significantly within a relatively short distance from the shock. At $B=5\times 10^{-5}$, the observed turnover frequency will
be approximately $10$\,GHz; the field should not be much larger than this value or the assumption that the optical depth is low
breaks down.

\section {Discussion}

A full chi-squared analysis of the difference between the data and models presented has not been undertaken 
and is probably not appropriate: what such an analysis would show is that, in detail, the model and data fit 
rather poorly. This is a consequence of the many assumptions made in the very simple model, e.g. the cylindrical
symmetry, which, the data shows, can only be approximate. Nevertheless,
the previous section has shown that, for a range of plausible parameters, the recollimation shock
model reproduces the principal features of the polarized intensity images in the core structure of 1803+784.
Furthermore, parameters can be found for which the model describes these features quantitatively: 
for most,
model and data parameters agree to within $2.5\sigma$. Given the extreme simplicity of the model and
the small number of free parameters, this appears to be an impressive level of agreement. The fractional polarization of 
the downstream (or `jet') 
component, $m_{jet}$, is the property for which agreement is worst; its model polarization is too low by over 
$3.5\sigma$. Allowing a larger value of $\eta_1$ does increase the fractional polarization of the `jet' component,
but this is at the expense of further depleting the kinetic energy of the downstream flow, and also increasing
the fractional polarization (and therefore the contribution to $\chi^2$) for the `core' component. The higher 
polarization
associated with the `jet' component may be due to some residual ordering of the magnetic field by the first shock
that survives until the second shock is reached, and which is not taken into account by the model. This could 
lead to a higher degree of order in the magnetic field as the plasma emerges from the second shock and therefore
a higher fractional polarization than the model allows. It is also possible that other effects not included in the
model of Fig.\,4, such as velocity 
shear, play a role ordering the magnetic field.

Fig.\,5 shows that the convolved total intensity 
distribution of the model is clearly extended toward the lower, upstream side of the peak, as shown by the 
spacing of the contours. The reverse is true in the data,
where an extension to the downstream side of the peak is a manifestation of the extended jet structure 
(which is not included in the model). The first panel of Fig.\,5 shows clearly that the upstream extension in 
the model is due to the first of the two conical shocks, which is larger and fainter than the the second. 
The most obvious conclusion to draw from this is that the assumption of uniform flow between the two shocks has
led to too great a degree of compression as the plasma passes between them. If the flow direction changes, 
deviating slightly away from the axis, the two conical shock waves would be more similar in projected size and 
emissivity, thereby largely eliminating the low brightness extension seen in the fourth panel of Fig\,5. 
As this emission is due to the parts of the shock structure that are farthest upstream (by quite a large margin, 
given the small cone angle), it may also be that this extended emission is
suppressed by opacity. 

A further source of disagreement is the precise orientation of the polarization rods. There is a broad level of
agreement between model and data: both have on-axis $E$ field parallel to the axis and a roughly radial pattern
elsewhere. However, off axis, the polarization angle values do not agree precisely, and this is not reflected in
the comparison between model and data presented in Section 3. The disagreement is clear from the
orientation of the polarization rods in the outer parts of the source on the line joining the two polarization
minima. In the data (Fig.\,1) these are almost perpendicular to the axis, whereas in the model (Fig.\,5, panel 5)
they make an angle with the axis close to $60^{\circ}$. 
Inspection of the second panel in Fig.\,5 shows that the polarization angles in the region
farthest from the axis are
determined largely by the first (and largest) of the two conical shocks in the region furthest upstream.
This is the same region responsible for the unwanted upstream extension in total intensity and therefore
the discrepancy may be also be reduced if the shocks are more equal in size, or if the upstream emission
is suppressed by opacity, as suggested above.

From this discussion it is clear that, while the recollimation shock model seems very promising as a means
of explaining the core polarization in 1803+784, input from a numerical magnetohydrodynamical (MHD) investigation
would be useful as a means to examine discrepancies between the model and the 
data. 

In order to obtain a representation of the source, a very small viewing angle was required. At larger viewing angles,
the double peaks on axis and the minima either side disappear, and the model produces a
much smaller range of polarization angles than is seen in Figs.\,1 and 5. Most of the sources in a sample of blazars
will have viewing angles smaller than or comparable to $1/\gamma$, where $\gamma$ is the Lorentz factor. In the
model of Fig.\,5, the Lorentz factors are $\gamma_2=7.8$ and $\gamma_3=6.45$, respectively. 
Hence the emission is characterized by a Lorentz factor $\gamma \simeq 7$. For these Lorentz factors, the required 
polarization structures
were only obtained for angles $\theta<2^{\circ} \simeq 0.035$\,rad. Hence it might be expected that only
a fraction of order $(0.035/(1/7))^2 \simeq 0.06$ of sources in the sample would exhibit the interesting structure 
shown here. This is consistent with the fact that, of the 15 sources included in the sample of Jorstad et al.
(2005), only one showed structure of this type.
In more poorly aligned sources, fans of polarization rods covering a smaller range of polarization
angles might be expected, and examples of this kind of structure were seen in the cores of the sources 0420-014
and 0528+134 which were also included in the Jorstad et al. (2005) sample.

If the interpretation presented here is correct, then it is interesting to compare the location of the 
recollimation shock, and hence of the radio core, with the core position in other sources. 
The core position has been estimated from component speeds and time lags between X-ray dips and the ejection
of radio components in at least two radio jet sources: 3C\,120 (Marscher et al., 2002) and 3C\,111 (Chaterjee et al., 2011). 
In those sources the core location is between $10^{5}$ and $10^{6}$ times the black hole Schwarzchild
radius, $R_s$. For 1803+784, a BL~Lac object, the black hole mass has been estimated to be in the region of $4\times 10^{8}M_\Sun$,
(Wang et al., 2004), though the scaling relation from which this is estimated involves a significant scatter (at least
half a decade). The black hole--core separation must be at least the height of one of the conical shocks. The projected
height can
be estimated roughly as about half the separation of the polarization peaks in Fig.\,1, i.e. about $0.1\,mas$. 
If the line of sight angle $\theta \simeq 1.5^{\circ}$ (as in the model of Table 1) the deprojected height is then about $30$\,pc.
These values give a black hole--core separation of approximately $3\times 10^{6}\,R_{s}$, somewhat larger
than for 3C\,120 or 3C\,111. However,  $\theta$ could be larger if the Lorentz factors are correspondingly
smaller, and the black hole mass is likely to be uncertain by at least a factor of three. Given these uncertainties, it is at least
possible that the black hole--core separation in units of $R_s$ is of the same order as that in the other sources.

A further feature evident from Fig.\,1 is the variation in direction of the symmetry axis between the two epochs 
shown. In fact, there is a definite
wobble in the direction of the axis from epoch to epoch, similar to that noted by Denn et al. (2000) and Stirling
et al. (2003) in BL\,Lac. These variations are too large to be caused by possible changes in orientation of the
nearly circular beam.
What is noteworthy here is how much more evident are these variations in polarization
for this distinctive type of structure than in total intensity. A detailed analysis of these variations 
is left for future work. 

\section{Conclusions.} 
Observations of a number of sources from the AGN sample studied by Jorstad
et al. (2005, 2007) at $43$\,GHz revealed unusual patterns of polarization
structure reminiscent of structures predicted by the Conical Shock models of Cawthorne \& Cobb, 
(1990) and Cawthorne (2006). The most interesting of these was found in the source 1803+784, 
which, at most epochs, has two on-axis polarization peaks and two polarization minima along 
a line perpendicular to the axis and lying between the two peaks. Such features were visible
in this source at most of the 17 observing epochs. Single conical shocks alone acting on plasma with
a largely disordered upstream magnetic field
cannot reproduce these properties, but the analytical simulations presented in this paper
show that twin conical shocks of the kind that occur in recollimation shocks can yield these 
characteristics for a range of reasonable model parameters, when the jet axis and the 
observer's line of sight are well aligned. The agreement between model and data is described
by defining five parameters that characterize the polarization image. For four of these, the
agreement is reasonable, but the fifth, the fractional polarization associated with downstream
polarized (or `jet') component,
is discrepant by $4.4\sigma$ (October 1998) and $3.6\sigma$ (October 1999). Some aspects of the total 
intensity and polarization
angle distributions are also in disagreement. The root of these disagreements seems to be
the simplifications that were made regarding
the progress of the emitting gas between the two conical shocks,
and this points to the need for a numerical (MHD) investigation in which such assumptions are not required. 

Despite these disagreements, the ability of such a simple model with a small number of free parameters
to reproduce the principal polarization features observed in 1803+784 seems impressive, and suggests that
the core polarization in 1803+784 may well be associated with a recollimation shock. This would appear
to be the most direct evidence so far for associating the core emission in an extragalactic radio source with
a structure of this kind. It strengthens the view that the core components in high frequency observations
represent the point at which the quiescent jet structure is first disturbed, and that this disturbance may
in many instances be due to recollimation shock structure.  

\section{Acknowledgments}

TVC thanks the Director of the Jeremiah Horrocks Institute at the University of Central 
Lancashire for a sabbatical that allowed completion of this work. He also thanks Dr. Denise
Gabuzda and the Physics Department at University College Cork for providing sanctuary in
which much of the work presented here was performed. Dr. Gabuzda also provided many useful
comments on the paper, as did Dr. Jose--Luis G\'{o}mez and the anonymous referee. 
Mr. Colm Coughlan assisted with some of 
the analysis presented here.   
The research at Boston University was supported in part by US National
Science Foundation grant AST-0907893.
The VLBA is an instrument of the National Radio Astronomy Observatory. The
National Radio Astronomy Observatory is a facility of the National Science
Foundation operated under cooperative agreement by Associated
Universities, Inc.
 
\begin{center}
REFERENCES. \\
\end{center}
\begin{flushleft}
Agudo, I., G\'{o}mez, J.-L., Mart\'{i}, J.-M, Ib\'{a}\~{n}ez, J.-M., Marscher, A.\,P.,
Alberdi, A., Aloy, M.-A., Hardee, P.\,E. 2001, ApJ, 549, L183 \\
Agudo, I. 2009, ASPC, 402, 300 \\
Agudo, I., G\'{o}mez, J.-L., Casadio, C., Cawthorne, T.\,V., Roca-Sogorb, M. 2012, ApJ,
752, 92 \\ 
Britzen, S., Kudryavtseva, N.\,A., Witzel, A., Campbell, R.\,M., Ros, E., Karouzos, M.,
Mehta, A., Aller, M.\,F., Aller, H.\,D., Beckert, T., Zensus, J.\,A. 2010, A\&A, 511, 57 \\
Canvin, J.\,R., Laing, R.\,A. 2004, MNRAS, 350, 1342 \\ 
Cawthorne, T.\,V., Cobb, W.\,K. 1990, ApJ, 350, 536 \\
Cawthorne, T.\,V. 2006, MNRAS, 367, 851 \\
Chatterjee, R., Marscher, A.\,P., Jorstad, S.\,G., Markowitz, A., Rivers, E., Rothschild, R.\,E., McHardy, I.\,M.,
Aller, M.\,F., Aller, H.\,D., Lahteenmaki, A., Tornikson, M., Harrison, B., Agudo, I., G\'{o}mez, J.-L., Taylor, B.\,W.,
Gurwell, M. 2011, ApJ, 743, 43 \\
Daly, R.\,A., Marscher, A.\,P. 1988, ApJ, 334, 539 \\ 
D'Arcangelo, F.\,D., Marscher, A.\,P., Jorstad, S.\,G., Smith, P.\,S., Larionov, V.\,M.,
Hagen-Thorn, V.\,A., Kopatskaya, E.\,N., Williams, G.\,G., Gear, W.\,K. 2007, ApJ, 659, L107 \\
D'Arcangelo, F.\,D. 2010, Ph.D. Thesis, Boston Univ. \\
Denn, G., Mutel, R., Marscher, A.\,P. 2000, ApJS, 129, 61 \\ 
Eilek, J.\,A., Hughes, P.\,A. 1991, Beams and Jets in Astrophysics, (Cambridge: CUP), 428\\  
Falle, S.\,A.\,E.\,G., Wilson, M.\,J. 1985, MNRAS, 216, 79 \\ 
Gabuzda, D.\,C., Sitko, M.\,L. 1994, AJ, 107, 884 \\ 
G\'{o}mez, J.-L., Mart\'{i}, J.\,M., Marscher, A.\,P., Ib\'{a}\~{n}ez, J.\,M., Marcaide, J.\,M.
1995, ApJ, 449, L19 \\
G\'{o}mez, J.-L., Mart\'{i}, J.\,M., Marscher, A.\,P., Ib\'{a}\~{n}ez, J.\,M., Alberdi, A. 1997,
ApJ, 482, L33 \\ 
Jorstad, S.\,G., Marscher, A.\,P., Lister, M.\,L., Stirling, A.\,M., Cawthorne, T.\,V., 
Gear, W.\,K., G\'{o}mez, J.-L., Stevens, J.\,A., Smith, P.\,S., Forster, J.\,R., Robson, E.\,I.
2005, AJ, 130, 1418 \\
Jorstad, S.\,G., Marscher, A.\,P., Stevens, J.\,A., Smith, P.\,S., Forster, J.\,R., 
Gear, W.\,K., Cawthorne, T.\,V.,  Lister, M.\,L., Stirling, A.\,M., G\'{o}mez, J.-L.,
Greaves, J.\,S., Robson, E.\,I. 2007, AJ, 134, 799 \\
Hovatta, T., Lister, M., Aller, M.\,F., Aller, H.\,D., Homan, D.\,C., Kovalev, Y.\,Y., Pushkarev, A.\,B.,
Savolainen, T. 2012, AJ, 144, 105 \\
Hughes, P.\,A. 2005, ApJ, 621, 635 \\
Lawrence, C.\,R., Zucker, J.\,R., Readhead, A.\,C.\,S., Unwin, S.\,C., Pearson, T.\,J., Xu W.
1996, ApJS, 107, 541 \\ 
Leahy, J.\,P. 1991, in Beams and Jets in Astrophysics, Ed. P. A. Hughes, CUP, p. 100\\ 
Lind, K.\,R., Blandford, R.\,D. 1985, ApJ, 295, 358 \\
Mahmud, M., Gabuzda, D.\,C., Bezrukovs, V. 2009, MNRAS, 400, 2 \\
Marscher, A.\,P., Jorstad, S.\,G., G\'{o}mez, J.-L., Aller, M.\,F., Terasantra, H., Lister, M., Stirling, A.
2002, Nature, 417, 625 \\
Marscher, A.\,P., Jorstad, S.\,G., D'Arcangelo, F.\,D., Smith, P.\,S., Williams, G.\,G., 
Larionov, V.\,M., Oh, H., Olmstead, A.\,R., Aller, M.\,F., Aller, H.\,D. and 13 coathors
2008, Nature, 212, 2002 \\
Nalewajko, K. 2009, MNRAS, 395, 524 \\
O'Sullivan, S.\,P., Gabuzda, D.\,C. 2009, MNRAS, 400, 26 \\
Papageorgiou, A., Cawthorne, T.\,V., Stirling, A.\,M., Gabuzda, D.\,C., Polatidis, A.\,G.
2006, MNRAS, 373, 449 \\
Roca-Sogorb, M., Perucho, M., G\'{o}mez, J.-L., Marti, J.\,M., Anton, L., Aloy, M.\,A., Agudo, I. 2008, in 
Extragalactic Jets: Theory and Observation from Radio to Gamma Ray, ASP Conference Series, 386, 488\\
Roca-Sogorb, M. 2011, PhD Thesis, University of Valencia \\
Roca-Sogorb, M., G\'{o}mez, J.-L., Agudo, I., Marscher, A.\,P., Jorstad, S.\,G. 2010, ApJ, 
712, 160 \\
Scheuer, P.\,A.\,G. 1987, in Astrophysical Jets and their Engines, Ed. W. Kundt, D. Reidel,
p. 137 \\
Stirling, A.\,M., Cawthorne, T.\,V., Stevens, J.\,A., Jorstad, S.\,G., Marscher, A.\,P.,
Lister, M.\,L., G\'{o}mez, J.-L., Smith, P.\,S., Agudo, I., Gabuzda, D.\,C., 
Robson, E.\,I., Gear, W.\,K. MNRAS, 341, 405 \\
Wang, J.--M., Luo, B., Ho, L.\,C. 2004, ApJ, 615, L9 \\
\end{flushleft}

\end{document}